\documentclass[prd,preprint,tightenlines,floatfix,showpacs,preprintnumbers,nofootinbib]{revtex4}
 \usepackage[dvips,final]{graphicx}
\begin{document}
\thispagestyle{empty}
%\date{\today}
\title{Riemann $\zeta(3)$-
terms   in  
 perturbative QED series, conformal symmetry  and the  analogies   with 
structures of  multiloop effects in  $N=4$ 
supersymmetric Yang-Mills theory}     
      \author{A.~L.~Kataev}
\email{kataev@ms2.inr.ac.ru}
\affiliation{Institute for Nuclear Research of the Academy of Sciences of Russia, 
117312  Moscow, Russia}
%\begin{center}

\begin{abstract}
 As was  discovered recently, the    5-loop perturbative quenched QED  approximation  
to  the QED $\beta$-function consist from the rational term     and the term   
 proportional to $\zeta(3)$-function. It is stressed,  that      
 this feature is  also   manifesting  itself in  
 the  conformal invariant pqQED series for the   
  4-loop  approximation  to the  anomalous mass dimension.
   The 4-loop pqQED  expression for  
 the singlet contribution into the   
 Ellis-Jaffe polarized sum rule  is obtained. It     
  coincides with the similar approximation  for the  non-singlet coefficient functions
 of the  Ellis-Jaffe sum rule and of the    Bjorken polarized   sum rule. 
 It is stressed  that  this property is valid  in all orders of perturbation theory 
  thanks to the  conformal symmetry of pqQED series and to the  
 Crewther relation, which  
 relates   
 non-singlet and singlet coefficient functions  of the   Ellis-Jaffe sum rule 
 with the coefficient functions of the  non-singlet and singlet  Adler D-functions.
 The basic steps of derivation of the  Crewther relation in the singlet channel  from  
 the  AVV triangle diagram are  outlined.   The similarities between analytical structures       
 of asymptotic  series for the     coefficient functions in pqQED  
 and   for  the  anomalous dimensions in  $N=4$ conformal invariant supersymmertic 
 Yang-Mills  theory are observed. The guess is proposed  that the appearance 
 of $\zeta(3)$-terms   in the pqQED expressions and the absence 
 of $\zeta(5)$-terms at the same level is the indication 
 of absence of "wrapping" interactions in pqQED.    
\end{abstract}
%\end{abstract}
\maketitle
%\newpage

%%%%%%%%%%%%%%%%%%%%%%%%%%%%%%%%%%%%%%%%%%%%%%%%%%%%%%%%%%%%%%%%%%%%%%%%%%%%%%
%%%%%%%%%%%%%%%%%%%%%%%%%%%%%%%%%%%%%%%%%%%%%%%%%%%%%%%%%%%%%%%%%%%%%%%%%%%%%%
{\bf 1.~Introduction} \\
\label{Sec:I}
%%%%%%%%%%%%%%%%%%%%%%%%%%%%%%%%%%%%%%%%%%%%%%%%%%%%%%%%%%%%%%%%%%%%%%%%%%%%%%

Among consequences  of the recent  advanced analytical QCD  
calculations of the   5-loop perturbative    corrections to the  
$e^+e^-$-annihilation Adler $D$-function \cite{BChK07}-\cite{BChK10} and to the 
Bjorken sum rule  of  polarized lepton-hadron deep-inelastic scattering 
\cite{BChK10} is the single-fermion 5-loop contribution 
to the QED renormalization-group $\beta$-function \cite{BChK07},\cite{BChK10}.  

Single-fermion loop approximation of perturbative QED is sometimes  associated 
with the term "quenched QED" which,  rigorously  speaking, is commonly accepted 
in non-perturbative QED studies (for the recent considerations see e.g.  
\cite{LQED},
\cite{VG}). In the case of "perturbative quenching" the related Feynman diagrams 
with internal photon vacuum polarization diagrams are not considered and therefore 
the coupling constant of QED  is renormalized  only by the subset  of   vacuum 
polarization subgraphs with one external fermion loop and 0$\leq {\rm n } \leq \infty$ internal 
photon lines. It can be shown, that within this approximation  the  QED vacuum 
polarization function and the related contribution  to the 
 QED $\beta$-function 
do not 
depend from the choice of renormalization scheme 
\cite{Broadhurst:1992za}. 
  
It should be stressed that  massless   
"perturbative quenched QED" (pqQED) obey  
the important property of  {\bf conformal symmetry}. Indeed, within this 
limit the   renormalized QED  coupling constant stay  fixed and is not running.  
Conformal symmetry    allows to   
connect    multiloop  expression for the massless 
pqQED part  of the  $e^+e^-$ 
annihilation  Adler D-function   with the pqQED contribution to   
the  Bjorken polarized sum rule \cite{Crewther:1972kn}. The details of 
this statement are explained in Ref. \cite{Kataev:2008sk}, where    
the way how to check   the  appearance of $\zeta(3)$ in  the 5-loop result 
of Ref.\cite{BChK07} by additional 5-loop perturbative QCD calculations  was 
outlined. 
This way  was followed in  Ref. \cite{BChK10}, where not only      
the  5-loop pqQED contribution 
to the Bjorken polarized sum rule, derived by  the back-of-envelope 
calculations     of  Ref.\cite{Kataev:2008sk}, 
were  reproduced, 
 but  the explicit 5-loop form of the   $\beta$- function 
factorizable generalization of the    Crewther relation in QCD  
was obtained as well \cite{BChK10}. This relation was  
 discovered previously  at the 4-loop level in  Ref.\cite{Broadhurst:1993ru}. 
 Its all-order
validity was  proved later in Ref. \cite{Crewther:1997ux}. Note, that in Ref. 
\cite{Kataev:2010dm}      
new   type  of the $\beta$-function factorizable QCD generalization of the 
 Crewther  relation was proposed.
Its more detailed study is in progress.

Study    of the origin of the 
appearance of the "puzzling  $\zeta(3)$-term"  in the 5-loop correction 
of the pqQED  $\beta$-function
is an  interesting  theoretical problem.   
Contributions of  Riemann $\zeta(3)$-functions  
were  already detected at the intermediate stages of the diagram-by-diagram 
calculations during the 3-loop calculations,  
which were  performed in the works of 
Refs.  \cite{Rosner}, \cite{ChKT1},\cite{ChKT2}. But  
after summing up  
all corresponding 3-loop graphs the terms, proportional to 
$\zeta(3)$, cancelled in the ultimate result. The details of these 
cancellations were followed up 
in Ref. \cite{ChKT1} and in Ref.\cite{Broadhurst:1995dq} later on.  
In the work of Ref. \cite{ChKT1} the explicit
diagrammatic analysis was made, while the analysis of  Ref.\cite{Broadhurst:1995dq} 
was performed   within the knot-theory formalism.
In Ref.\cite{BKZ} the guess was formulated   that the rationality of  
the 3-loop level expression  is related  to  really 
existing property of the  conformal symmetry 
of the pqQED  series. Using this guess and neglecting 
integrals, which generate $\zeta(3)$-terms at the 3-loop level,  the authors 
of Ref.\cite{BKZ} reproduced the  original result of Ref.\cite{Rosner}.

Analytical calculations of the 4-loop   QED $\beta$-function in the class of 
modified  
minimal subtractions  renormalization scheme (and of the  $\overline{MS}$-scheme 
among them) 
were  finally  completed  
in Ref.\cite{Gorishnii:1990kd}. In these calculations 
in addition to $\zeta(3)$-term, the terms proportional to $\zeta(5)$
appeared as well, but both types of transcendental functions disappeared 
in the expression for the 4-loop correction to the $\beta$-function 
in the pqQED approximation.   
The absence of the  $\zeta$-functions   at the 4-loop level    
were confirmed in Ref.\cite{Broadhurst:1999zi} using the Schwinger-Dyson equations, 
where the rational  4-loop expression  of  $O(A^4)$ contribution to 
the pqQED  $\beta$-function,  obtained 
in Ref.\cite{Gorishnii:1990kd}, was reproduced as well. The 
 background field method calculations of Ref. \cite{Broadhurst:1999xk}
clarified the origin  of rationality of this order $A^4$ term. 

In view of the 3 and 4-loop cancellations of the  
transcendental Riemann functions at the 3- and 4-loop levels,  
the explicit manifestation of the  $\zeta(3)$-function  at the 5-loop level   
 was  considered   as a puzzle. 
 In this paper we will 
show  that this is not the puzzle at all, but the regular feature,  which 
is consistent with the structure  of perturbative series in another conformal 
invariant model, namely $N=4$ supersymmetric (SUSY) Yang-Mills theory. \\   

%%%%%%%%%%%%%%%%%%%%%%%%%%%%%%%%%%%%%%%%%%%%%%%%%%%%%%%%%%%%%%%%%%%%%%%%%%%%%%
%%%%%%%%%%%%%%%%%%%%%%%%%%%%%%%%%%%%%%%%%%%%%%%%%%%%%%%%%%%%%%%%%%%%%%%%%%%%%%
{\bf 2.~Manifestations  of  $\zeta(3)$ functions in  perturbative quenched 
QED series   and its cancellation in the 4-loop expression for Ellis-Jaffe 
sum rule.} \\
\label{Sec:II}
%%%%%%%%%%%%%%%%%%%%%%%%%%%%%%%%%%%%%%%%%%%%%%%%%%%%%%%%%%%%%%%%%%%%%%%%%%%%%%

It is important to have a look, how in perturbative quenched QED series 
$\zeta(3)$-terms are manifesting themselves. Consider first the original 
result of Refs.\cite{BChK07},\cite{BChK10} for the 5-loop expression  
of the pqQED $\beta$-function, namely 
\begin{eqnarray}
\label{beta1}
\beta_{QED}^{[1]}(A)&=&\frac{4}{3}A+4A^2-2A^3-46A^4+\bigg(\frac{4157}{6}+
128\zeta(3)\bigg)A^5 + O(A^6) \\
&=& \frac{4}{3}A\times C_{D}^{ns}(A)~~~~.
\end{eqnarray} 
The  coefficient function  is defined from the QCD expression for  
the non-singlet contribution to the $e^+e^-$-annihilation Adler D-function
\begin{equation}
D^{ns}(A_s)=3\sum_{F}Q_F^2C_D^{ns}(A_s)~~~.
\end{equation}
The QED and QCD perturbation-theory  expansion parameters are normalized  as 
$A=\alpha/(4\pi)$ and  $A_s=\alpha_s/(4\pi)$ 
with  $\alpha$ and $\alpha_s$ being  the renormalized QED and QCD coupling constants. 

It is interesting to have a look whether in perturbative quenched QED 
there are any other  
renormalization group function for the gauge-invariant operators, 
which contain $\zeta(3)$-function in high order corrections.

Consider first  perturbative series for the  
anomalous mass  dimension in pqQED. Its expression differs  
from the anomalous dimension  
 of the operator $\overline{\Psi}\Psi$ by overall sign only, and therefore, 
for the reason of rigour it is better not to introduce mass term 
in the QED lagrangian, and consider massless  conformal invariant limit 
of the QED series   for the anomalous dimension function 
 $\gamma_{\overline{\Psi}\Psi}(A)=-\gamma_{m}(A)$.
Its expression can be obtained from the 4-loop QCD calculations of   
the mass  anomalous dimension function $\gamma_{m}(\alpha_s)$, performed in  
Ref.\cite{Vermaseren:1997fq} and in Ref.\cite{Chetyrkin:1997dh} independently.
It is more convenient to  use the results of \cite{Vermaseren:1997fq}, since 
this 
work contains  the 
explicit dependence of the 4-loop expression for  $\gamma_{m}(\alpha_s)$ from 
 Casimir operators ${\rm C_F}$, ${\rm C_A}$, normalization factor  ${\rm T_F}$ 
 and   the number of quarks flavours ${\rm N_F}$.
The choice 
${\rm C_F=1}$, ${\rm C_A=0}$, ${\rm 
T_F=1}$ and ${\rm N_F=0}$ corresponds to the case 
of pqQED approximation. The pqQED expression for the anomalous 
dimension of the gauge-invariant operator $\overline{\Psi}\Psi$ 
has the following form 
\begin{equation}
\label{gamma}
\gamma_{\overline{\Psi}\Psi}^{pqQED}(A)= -3A-\frac{3}{2}A^2-\frac{129}{2}A^3+
\bigg(\frac{1261}{8}+336\zeta(3)\bigg)A^4+O(A^5)
\end{equation}

The analytical structure  of this series was    
already investigated  in 
Ref.\cite{Broadhurst:1999zi} using the Shwinger-Dyson approach.   
In view of the appearance of $\zeta(3)$-term 
in the pqQED part  for the QED $\beta$-function (see Eq.(\ref{beta1})),  
it is worth to attract more attention 
to the appearance of $\zeta(3)$-term in the 4-loop correction in   
Eq.(\ref{gamma}).  Moreover, the  4-loop manifestation 
of $\zeta(3)$-term in the conformal 
invariant  expression
of Eq.(\ref{gamma}) indicate that the similar feature may 
manifest itself  in   other 
pqQED series as well.   The anticipating its  manifestation 
cancellations of $\zeta(3)$-terms at the intermediate stages of lower order 
calculation  {\bf should also hold} 
in the    series of Eq.(\ref{gamma}). This  statement is the consequence 
of  the experience 
gained in the process of evaluation of 3-loop counter-terms in QCD during the 4-loop  
calculations,
which result in the publications of the works of  Refs. \cite{Gorishnii:1990kd},
\cite{Gorishnii:1990vf}, \cite{Gorishnii:1991hw}. 

Note, that the expression for Eq.(\ref{gamma}) follows from the calculations of 
the renormalization group function of "vertex operator". 
In the  case of  calculations of renormalization-group quantities, related to 
two-point functions,  $\zeta(3)$-term  is appearing one loop 
later, namely at the 5-loop order (see Eq.(\ref{beta1})). It enters the    
expressions for the non-singlet coefficient functions of the 5-loop  
$O(A^4)$-corrections to 
the $e^+e^-$-annihilation Adler $D$-function 
and the   Bjorken polarized deep-inelastic scattering sum rule, defined in QCD 
as  
\begin{equation}
\label{Bjp}
Bjp(Q^2)=\int_0^1\bigg[g_1^{lp}(x,Q^2)-g_1^{ln}(x,Q^2)\bigg]dx=
\frac{1}{6}g_aC_{Bjp}^{ns}(A_s(Q^2))~~~.
\end{equation}
Indeed, comparing Eq.(1) with Eq.(2), one can get:  
\begin{equation}
\label{D}
C_D^{ns}= 1+3A-\frac{3}{2}A^2-\frac{69}{2}A^3+\bigg(\frac{4157}{8}+
96\zeta(3)\bigg)A^4 +O(A^5)~~~.
\end{equation}
The similar 5-loop expression for the coefficient function of the Bjorken sum rule, 
given in Ref. \cite{Kataev:2008sk} and confirmed by diagram-by-diagram 
calculations  in Ref. \cite{BChK10},  reads:
\begin{equation}
\label{Bj}
C_{Bjp}^{ns}=1-3A+\frac{21}{2}A^2-\frac{3}{2}A^3-\bigg(\frac{4823}{8}
+96\zeta(3)\bigg)A^4 +O(A^5)~~~.
\end{equation}
These quantities do not contain anomalous dimension terms.

The  logic of the discussions presented above 
leads to the  conclusion  that  in  the pqQED series for the quantities, 
which are related with  the   
non-zero anomalous renormalization constant, $\zeta(3)$-should cancel down   
1-loop prior their  manifestation in Eq.(\ref{D}) and Eq.(\ref{Bj}), namely 
at the level of  $O(A^3)$-corrections. 

To  verify this statement consider now the   
Ellis-Jaffe sum rule  
of the deep-inelastic scattering of   polarized leptons on protons. 
In QCD it is defined as  
\begin{equation}
\label{EJ}
EJp(Q^2)=\int_0^1 g_1^{lp}(x,Q^2)dx =  C_{Bjp}^{ns}(A_s(Q^2))\bigg(\frac{1}{12} a_3+
\frac{1}{36}a_8\bigg) 
+ C_{EJp}^{s}(Q^2) \frac{1}{9}\Delta\Sigma(Q^2)
\end{equation}
where $a_3=\Delta u-\Delta d$, $a_8=\Delta u+\Delta d-2\Delta s$,
$\Delta u$, $\Delta d$ and $\Delta s$ are the polarized distributions and 
$\Delta\Sigma$ depends from the scheme choice. In the $\overline{MS}$-scheme 
it is defined as $\Delta\Sigma=\Delta u+\Delta d+\Delta s$, while in the 
Adler-Bardeen scheme it contains the additional additive contribution 
from  polarized gluon distribution $\Delta G$.

The 4-loop QCD  corrections to the  coefficient function of the singlet part of 
Ellis-Jaffe sum rule were calculated in Ref. \cite{Larin:1997qq} using the method 
of the dimensional regularization. In the framework of the 
dimensional regularization the final expression for the    
singlet coefficient function can be presented as the ratio of two 
functions \cite{Larin:1997qq}:  
\begin{equation}
\label{C}
C_{EJp}^{s}=\overline{C}_{EJp}^{s}/Z_5^s~~~.
\end{equation}

Here $Z_5^s$ is the finite singlet renormalization constant  
of the operator 
$\overline{\Psi}\gamma_{\mu}\gamma_5\Psi$, which should be calculated 
within   dimensional regularization and the $\overline{MS}$-scheme.   
This finite  constant is similar to the   
finite constant  $Z_5^{ns}$ in the definition  of  
the non-singlet axial operator 
$\overline{\Psi}\gamma_{\mu}\gamma_{5}(\lambda^a/2)\Psi$ within dimensional 
regularization. It   enters in the 
procedure of calculations of high order QCD corrections  
to the Bjorken polarized sum rule at the 3-loop \cite{Gorishnii:1985xm}, \cite{Zijlstra:1992kj}
and 4-loop \cite{Larin:1991tj} levels.  
In view of the property, that the expression for $Z_5$ differs from 
$Z_5^{ns}$ by the corrections, which come from the light-by-light-type 
scattering graphs  \cite{Larin:1993tq}, in the pqQED 
limit these constants coincide. Therefore, the   4-loop corrections 
in  Eq.(\ref{C}) are determined by the ratio of 
the following pqQED expressions for the coefficient function   
\begin{equation}
\label{Cs}
\overline{C}_{EJ}^{s}=1-7A+\frac{89}{2}A^2-\bigg(\frac{1397}{6}-96\zeta(3)\bigg)A^3+
O(A^4)~~~.
\end{equation}
and for the finite renormalization constants, namely
\begin{equation}
\label{Z5}
Z_5^{s}|_{pqQED}= Z_5|_{pqQED}= 1-4A+22 A^2 + 
\bigg(-\frac{370}{3}+96\zeta_3\bigg)A^3+O(A^4)~~~.  
\end{equation} 
The expressions   of Eq.(\ref{Cs}) and of Eq.(\ref{Z5}) are extracted  
from the results of calculations  of Ref. \cite{Larin:1997qq} and 
Ref. \cite{Larin:1991tj} correspondingly.
Notice  the appearances of $\zeta(3)$-terms in the coefficients 
of the $O(A^3)$-corrections to  Eq.(\ref{Cs}) and Eq.(\ref{Z5}). However, these 
terms cancel each other in our new   ultimate 4-loop  pqQED result   
for the coefficient of  order $A^3$ approximation  to the singlet coefficient 
function: 
\begin{equation}
\label{SI}
C_{EJp}^{s}(A)= 1-3A+\frac{21}{2}A^2-\frac{3}{2}A^3+O(A^4)
\end{equation}
and {\bf coincide} with the similar expression for the pqQED series 
of Eq.(\ref{Bj}).  

At the  possible next   step of analytical calculations of Eq.(\ref{Cs}) $\zeta(5)$ 
must manifest itself. Indeed, not presented yet   
next term in the result of  Eq.(\ref{Z5}) for the renormalization constant 
$Z_5|_{pqQED}$,   
evaluated during the calculations  of 5-loop perturbative  
corrections to the Bjorken polarized sum rule \cite{BChK10}, must  contain 
$\zeta(5)$ 
function, while corresponding $\zeta(7)$-terms should cancel in the  
expressions for its $O(A^3)$-corrections.   However, $\zeta(3)$ should 
remain in the expression of the coefficient of the   $O(A^4)$-correction to 
Eq.(\ref{SI}), since the following identity  
\begin{equation} 
\label{SEB}
C_{EJp}^{s}(A)=C_{Bjp}^{ns}(A)
\end{equation}
holds in pqQED in {\bf all orders of perturbation theory} and is the 
consequence of the  of the axial variant of  Crewther relation.  
 
 Let me outline the basic steps of the proof of this statement  in the 
 momentum space. These steps were  first discussed in   
 Ref. \cite{Kataev:1996ce} together with more detailed proof of the original 
 non-singlet Crewther relation in the momentum space \cite{Gabadadze:1995ei}\footnote{The  
  Crewther  relation  \cite{Crewther:1972kn} was originally derived  in the coordinate 
space.}. 

The proof is based on the application of the operator product expansion 
approach  to  the   3-point function with the axial
singlet current :
\begin{equation}
T_{\mu\alpha\beta}^{ab}(p,q)=i\int<0|TA_{\mu}(y)
V_{\alpha}^a(x)V_{\beta}^b(0)|0>e^{ipx+iqy}dxdy
\label{avv}
\end{equation}
where $A_{\mu}=\overline{\psi}\gamma_{\mu}\gamma_{5}\psi$.
Keeping the singlet structure in the operator-product expansion of
the two non-singlet vector currents, one can get
\begin{equation}
i\int T V_{\alpha}^aV_{\beta}^be^{ipx}dx|_{p^2\rightarrow\infty}
\approx C_{\alpha\beta\rho}^{SI,ab}A_{\rho}(0)+other~ structures~~~.
\label{vv}
\end{equation}
where 
\begin{equation}
C_{\mu\nu\alpha}^{SI,ab}\sim i\delta^{ab}\epsilon_{\mu\nu\alpha\beta}
\frac{q^{\beta}}{Q^2}C_{EJp}^{s}(a_s)~~~~
\label{csi}
\end{equation}
The second ingredient  in the  singlet version of the Crewther
relation appears after consideration of vacuum expectation value 
of the product of two axial currents 
\begin{equation}
i\int<0|T A_{\mu}(x) A_{\nu}(0)|0>e^{iqx}dx=\Pi_{\mu\nu}^{SI}(q^2)~~~.
\label{ax3}
\end{equation}
In this channel one can also  define Adler function and 
its coefficient function $C_D^{s}(A_s)$ as well. 
Taking now the conformal symmetry limit,  
it is possible to get 
the singlet variant of the Crewther relation \cite{Kataev:1996ce}, namely  
\begin{equation}
\label{EJD}
C_{EJp}^{s}(A)\times C_D^{s}(A)|_{conf-sym }=1~~~.
\end{equation}
This expression should be compared with 
the similar expression for the non-singlet  Crewther 
relation \cite{Gabadadze:1995ei},  which reads 
\begin{equation}
\label{BJD}
C_{Bjp}^{ns}(A)\times C_D^{s}(A)|_{conf-sym }=1~~~.
\end{equation}
In the massless pqQED approximation   the following identity takes place   
\begin{equation}
\label{Si}
C_{D}^{s}(A)=C_D^{ns}(A)~~~.
\end{equation}
Comparing now  Eq.(\ref{EJD}) with Eq. (\ref{BJD}) and taking into 
account Eq.(\ref{Si}), 
I  find that   the expression of Eq.(\ref{SEB}) is indeed 
valid in  all orders of perturbation theory. 

Note once more that pqQED is the  {\bf conformal invariant 
version} of massless  perturbative QED. Therefore, in order to understand deeper 
the status and nature of the manifestation of  odd $\zeta$-functions,  
it is important to have a look to  the structure  of perturbative series      
in some   other {\bf conformal invariant} 
theory  and   {\bf $N=4$ SUSY Yang-Mills theory} in particular. \\
%%%%%%%%%%%%%%%%%%%%%%%%%%%%%%%%%%%%%%%%%%%%%%%%%%%%%%%%%%%%%%%%%%%%%%%%%%%%%
%%%%%%%%%%%%%%%%%%%%%%%%%%%%%%%%%%%%%%%%%%%%%%%%%%%%%%%%%%%%%%%%%%%%%%%%%%%%%%
{\bf 3. Analytic structure for the  anomalous dimension of the  
Konishi operator in   $N$=4 SUSY Yang-Mills theory.} \\
\label{Sec:III}
%%%%%%%%%%%%%%%%%%%%%%%%%%%%%%%%%%%%%%%%%%%%%%%%%%%%%%%%%%%%%%%%%%%%%%%%%%%%%%
To demonstrate  that the explicit manifestation  of transcendental $\zeta(3)$-terms 
in high order perturbation theory corrections to renormalization group quantities 
does not contradict  {\bf conformal symmetry}  let us turn to the behaviour 
of perturbative series for the  anomalous dimensions in the massless 
 $N=4$ SUSY  Yang-Mills theory. 
Since its renormalization group  $\beta$-function
is identically equal to zero, this  theory possess the property 
of explicit conformal symmetry. The validity of this property at the 3-loop level 
was  discovered in Ref. \cite{Avdeev:1980bh}  by perturbative methods.  
Soon aftewards the absence of renormalization of the  coupling 
constant in this theory was proved within the  light-cone quantization approach 
\cite{Mandelstam:1982cb}. 

The absence of the  coupling constant  renormalization does not
mean  that there are no ultraviolet divergencies in the  massless $N=4$  
 Yang-Mills theory. Indeed,   calculations of anomalous 
dimensions of various operators in  this quantum field theory give non-zero results 
(see e.g. Refs. \cite{Kotikov:2002ab}- \cite{Velizhanin:2010cm}). 

Among the most interesting  are the ones, related to analytical  evaluation  of  
the anomalous dimension of the Konishi operator in high levels  of perturbation theory. 
The operator is defined as 
\begin{equation} 
O_{K}=tr \overline{\Phi}_i \Phi^i
\end{equation}
where $\Phi^i$ is the complex adjoint scalar field.     
The  expression  for the  anomalous dimension of this operator obey 
the interesting  property, namely the  transcendental functions 
$\zeta(3)$ and $\zeta(5)$ are manifesting themselves starting 
from the  {\bf 4-loop} perturbative corrections. Indeed, 
the direct   quantum field theory perturbative calculation, 
performed in terms of Feynman diagrammatic technique 
\cite{Velizhanin:2008jd}, gave the following result    
\begin{equation}
\label{Konishi}
\gamma_{K}(\lambda) = 12\lambda-48\lambda^2+336\lambda^3-
\lambda^4\bigg(2496-576\zeta(3)+1440\zeta(5)\bigg)+O(\lambda^5)
\end{equation}
where $\lambda=g^2N_c/(4\pi)^2$ and $N_c$ is the "number of colours" 
of $SU(N_c)$ gauge group. 
Note, that in $N=4$ SUSY Yang Mills gauge theory the values of Casimir
operators   are fixed as    
${\rm C_F}$=${\rm C_A}$=${\rm T_FN_F}$. Another interesting feature of 
$N=4$ SUSY Yang-Mills theory is that the property of   ${\rm AdS/CFT}$ correspondence 
\cite{Maldacena:1997re}-\cite{Witten:1998qj} links   
$N=4$ SUSY  Yang-Mills with the 
theory of  superstings in  $AdS_5\times S^5$. 

This property  opens the second way for the calculations of 
 anomalous dimensions in  $N=4$ SUSY Yang-Mills theory using 
quantum field theory of the superstring in $AdS_5\times S^5$ and taking into 
account its integrability property. This was done in Ref.\cite{Bajnok:2008bm}, 
where the coefficients of the series in Eq.(\ref{Konishi}) were calculated 
prior the work of Ref. \cite{Velizhanin:2008jd}.

This calculation is based  on the application of the Bethe Anzatz quantization.
Note, that using this ansatz it is possible to separate 
pure weak-coupling contribution from the one,  which interpolates between 
strong and weak coupling  \cite{Ambjorn:2005wa} and  is  responsible 
for the contribution of the L\"ucher corrections \cite{Luscher:1985dn}.  

In other words, its  
application allowed to demonstrate that  at  the level of  
{\bf order $\lambda^4$}    extra contributions,  
which describe  "wrapping effects"  of L\"ucher
corrections  \cite{Luscher:1985dn}, are manifesting themselves.  These effects are  
detectable both   at strong 
coupling constant regime  (see e.g. Ref. \cite{Arutyunov:2004vx})  and  weak 
coupling constant  regime   \cite{Fiamberti:2008sh}.

Perturbation-theory oriented clarification of these words is   encoded 
in the results of Ref. \cite{Bajnok:2008bm}. Indeed, the 4-loop
expression for $\gamma_K$ can be decomposed into two terms, namely  
\begin{equation}
\label{sum}
\gamma_{K}=\gamma_{asymp}(\lambda) + \gamma_{wr}(\lambda) 
\end{equation}
where 
\begin{eqnarray}
\label{as}
\gamma_{asymp}&=& 12\lambda-48\lambda^2+
336\lambda^3-\bigg(2820+288\zeta(3)\bigg)\lambda^4 +O(\lambda^5) 
 \\ \label{wr}
\gamma_{wr}(\lambda)&=&\bigg(324+864\zeta(3)-
1440\zeta(5)\bigg)\lambda^4 +O(\lambda^5)
\end{eqnarray}
The result of Eq.(\ref{as}) was first obtained  in 
Ref. \cite{Kotikov:2007cy}. 
The analytical calculations  of  overall  order 
$\lambda^4$-contribution 
and of its two parts are in agreement with the   calculations performed 
with     superspace diagrammatic  formalism \cite{Fiamberti:2008sh}. 
In its turn, the total expression  for the order 
$\lambda^4$-approximation of Eq.(\ref{sum}), obtained in Ref.\cite{Bajnok:2008bm}
and Ref. \cite{Fiamberti:2008sh}, coincide with the result 
of Eq.(\ref{Konishi}), obtained in Ref. \cite{Velizhanin:2008jd}
from direct Feynman diagrams calculations.  

This independent  calculation gave  real confidence 
in the correctness of final analytical expression and in  the  fact
that the {\bf asymptotic} part of 4-loop result for $\gamma_K$ (see Eq.(\ref{as}))
{\bf does  not contain} ${\bf \zeta(5)}$-contribution, which, 
together with additional 
pure rational  and $\zeta(3)$-contributions,  enter into 4-loop 
"wrapping" effects (for the diagrammatic explanation 
of the appearance of $\zeta(5)$ in Eq.(\ref{as}) see Ref. \cite{Velizhanin:2008pc})

The results of Eq.(24) should be compared with the pqQED ones, given in Sec.2. 
Compared with each other  they indicate, that 
Riemann $\zeta(3)$-puzzle is not the puzzle, 
but the regular 
feature of the asymptotic series in the conformal-invariant theories. Following 
this conclusion, one  should expect manifestation of $\zeta(3)$ and $\zeta(5)$ terms 
in the next-to-presented above  coefficients of the corresponding asymptotic 
perturbative series in the conformal invariant theories. This feature is realized 
in the results of calculations of   5-loop 
corrections  to the anomalous dimension of Eq.(\ref{as})  
in  $N=4$ SUSY Yang-Mills 
theory  \cite{Bajnok:2009vm}-\cite{Beccaria:2009rw}. Note that  
 $\zeta(7)$-terms are   
appearing in the 5-loop  "wrapping contributions" only
(see e.g. \cite{Fiamberti:2009jw},
 \cite{Lukowski:2009ce}).
Moreover,  $\zeta$-functions counting rule,  namely the appearance of extra 
$\zeta$-functions in high order wrapping contributions, 
is supported  by the results of six loop calculations   
(see Ref. \cite{Belitsky:2009mu} and 
Ref. \cite{Velizhanin:2010cm} ), which demonstrate the appearance 
of $\zeta(9)$-terms. \\

%%%%%%%%%%%%%%%%%%%%%%%%%%%%%%%%%%%%%%%%%%%%%%%%%%%%%%%%%%%%%%%%%%%%%%%
%%%%%%%%%%%%%%%%%%%%%%%%%%%%%%%%%%%%%%%%%%%%%%%%%%%%%%%%%%%%%%%%%%%%%%%
{\bf Conclusions.} \\
\label{sec:Concl}
%%%%%%%%%%%%%%%%%%%%%%%%%%%%%%%%%%%%%%%%%%%%%%%%%%%%%%%%%%%%%%%%%%%%%%%
%%%%%%%%%%%%%%%%%%%%%%%%%%%%%%%%%%%%%%%%%%%%%%%%%%%%%%%%%%%%%%%%%%%%%%%
In this work we introduce the way  of   
explaining the structure of analytical expression for high order    
corrections in  asymptotic perturbative series 
for the anomalous dimensions and coefficient functions of gauge-invariant 
operators in pqQED. The arguments, presented in this 
work, are useful for  realizing  
that  the appearance of $\zeta(3)$-terms in 
the pqQED series is rather   regular feature. 
This feature is  supported by the property of conformal symmetry.  

Indeed, the $\zeta$-functions  counting rules are also satisfied 
for dealing with  coefficients of the 
{\bf asymptotic}  perturbative series for the   
anomalous dimensions of  operators in super-conformal  $N=4$ SUSY
Yang-Mills gauge theory in the case when "wrapping interactions" are not  taken 
into account. These interactions are   responsible 
for the interpolation into the  regime  of large values 
of coupling constant.  

At present I do not know whether it is possible to find the signals of 
the existence of these interactions in the  strong-coupling phase  
of   quenched QED.  In the  case if 
 these interactions do exist, 
they may signal about  themselves   through the explicit manifestations of 
higher transcendentalities,   and $\zeta(5)$ in particular.  \\ 

%%%%%%%%%%%%%%%%%%%%%%%%%%%%%%%%%%%%%%%%%%%%%%%%%%%%%%%%%%%%%%%%%%%%%%%%%%%%%%%%%%
%%%%%%%%%%%%%%%%%%%%%%%%%%%%%%%%%%%%%%%%%%%%%%%%%%%%%%%%%%%%%%%%%%%%%%%%%%%%%%%%%%%%
{\bf Acknowledgements} 
%%%%%%%%%%%%%%%%%%%%%%%%%%%%%%%%%%%%%%%%%%%%%%%%%%%%%%%%%%%%%%%%%%%%%%%%%%%%%%%%%%%%%
%%%%%%%%%%%%%%%%%%%%%%%%%%%%%%%%%%%%%%%%%%%%%%%%%%%%%%%%%%%%%%%%%%%%%%%%%%%%%%%%%%%%%

It is the pleasure to thank D.J.  Broadhurst for the ideological support, J.L. Rosner 
for useful correspondence,    V.~N.~Velizhanin for  attracting my interest  
to the  studies, made in the area of perturbative analytic calculations in 
$N=4$ SUSY Yang-Mills theory. I wish to thank A.V. Kotikov   and C. Sieg for 
constructive  comments. 
   
I am grateful to L.N. Lipatov and  V.A. Matveev for the discussions of the definite 
scientific problems, studied  in this work.
  
This work  is done within the programs of research of   
the Scientific Schools of Russia   grant NSh-5525.2010.2 
and Russian Foundation of Basic Reaserch  grants  RFFI  
No. 08-01-00686, 08-02-01-01184.

%%%%%%%%%%%%%%%%%%%%%%%%%%%%%%%%%%%%%%%%%%%%%%%%%%%%%%%%%%%%%%%%%%%%%%%%%%%%%%%%%%%

\end{document}